\documentclass[namedreferences]{SolarPhysics}
\usepackage[optionalrh]{spr-sola-addons} 

\usepackage{graphicx}        
\usepackage{amssymb}         
\usepackage{url}             

\newcommand{\etal}{\textit{et al.}}

\usepackage{epstopdf}
\usepackage{amsmath}

\newcommand{\ka}{\kappa}

\newcommand{\sectt}{\sec^2\theta}

\newcommand{\costt}{\cos^2\theta}
\newcommand{\sint}{\sin\theta}

\newcommand{\half}{{\textstyle\frac{1}{2}}}
\newcommand{\threeontwo}{{\textstyle\frac{3}{2}}}

\newcommand{\re}{\mathop{\rm Re}\nolimits}

\newcommand{\rd}{\mathrm{d}}

\renewcommand{\k}{\mathbf{k}}

\newcommand{\F}{{{{}_2F_3}}}

\newcommand{\kperp}{k_{\!\perp}}
\newcommand{\kappaperp}{\kappa_{\!\perp}}
\newcommand{\kpar}{k_{\scriptscriptstyle\parallel}}

\newcommand{\bv}{Brunt-V\"ais\"al\"a}

\newcommand{\ri}{\mathrm{i}}

\newcommand{\ts}{\textstyle}

\newcommand{\calD}{{\mathcal{D}}}

\newcommand{\kp}{\kappa_0}
\newcommand{\kpz}{\kappa_z}

  \renewcommand{\ge}{\geqslant}

\newcommand{\aap}{\textit{Astron. Astrophys.}}

\newcommand{\apj}{\textit{Astrophys. J.}}
\newcommand{\apjl}{\textit{Astrophys. J. Lett.}}

\newcommand{\mnras}{\textit{Mon. Not. Roy. Astron. Soc.}}

\newcommand{\solphys}{\textit{Solar Phys.}}

\begin{document}
\begin{opening}

\title{An Exact Test of Generalized Ray Theory in Local Helioseismology}

\author{Shelley C. \surname{Hansen}$^{1}$\sep Paul S. \surname{Cally}$^{2}$}
\runningauthor{S.C. Hansen and P.S. Cally}

\runningtitle{Generalized Ray Theory in Local Helioseismology}

\institute{$^{1}$ 
School of Mathematical Sciences, Monash University,\\
Victoria 3800, Australia\\
(e-mail: \url{shelley.hansen@sci.monash.edu.au})\\ 
              $^{2}$ 
Centre for Stellar and Planetary Astrophysics,\\
School of Mathematical Sciences, Monash University,
Victoria 3800, Australia\\
(e-mail: \url{paul.cally@sci.monash.edu.au})\\  
             }

\begin{abstract}
Generalized Ray Theory (GRT) provides a simple description of MHD mode transmission and conversion between magneto\-acoustic fast and slow waves and is directly applicable to solar active regions. Here it is tested in a simple two-dimensional, isothermal, gravitationally-stratified model with inclined magnetic field using previously published exact solutions and found to perform very well.
\end{abstract}



\end{opening}

\section{Introduction}
Magneto\-hydro\-dynamic (MHD) mode conversion has long been proffered as an explanation of observations \cite{bdl88,braun95} of $f$- and $p$-mode ``absorption'' by sunspots. Several methods have been used to explore this question, including numerical solution of a differential eigenvalue problem
(\emph{e.g.}, \opencite{cbz}, \opencite{bc97}; \opencite{cc03}, \citeyear{cc05}; with results directly compared to the solar data in \opencite{ccb03} and \opencite{cccbd05}), and direct numerical simulation \cite{cb97,cally00,kc06,cgd08}. 

However, a very different technique, based on a generalization of ray theory \cite{cally06,sc06} presents several novel insights which greatly enhance interpretation and understanding. To date, this has been applied only in two dimensions (2D), in the sense that gravity, the magnetic field, and the direction of wave propagation all lie in the same plane, and we shall maintain that restriction here. This decouples the Alfv\'en wave from the problem, leaving only the fast and slow magneto\-acoustic waves (although see \opencite{cg08} for a quantitative estimation of 3D coupling).

In particular, Generalized Ray Theory (GRT) has verified that mode transmission/conversion occurs at or near the equipartition level ($z_\mathrm{eq}$) where the sound and Alfv\'en speeds coincide, $c=a$. A magneto\-acoustic wave incident on this level is partially transmitted (transmission coefficient $T$) and partially converted (coefficient $C$). GRT introduces the concept of the attack angle ($\alpha$), the angle between the wavevector and the magnetic field at the mode conversion level. If $\alpha$ is small, transmission dominates, but if it is large, then conversion is favoured.

In any plausible sunspot model, the sound speed ($c$) varies slowly with height ($z$) through the surface layers, but the Alfv\'en speed ($a=B/\sqrt{\mu\rho}$) increases rapidly due to the decreasing density ($\rho$). Here, $\mu$ is the magnetic permeability ($4\pi\times10^{-7}$ in SI units). For $z\ll z_\mathrm{eq}$ we have $c\gg a$, whilst $a\gg c$ in the opposite regime $z\gg z_\mathrm{eq}$.
In $c\gg a$ the fast wave is predominantly acoustic in nature, and the slow wave is largely magnetic. On the other hand, where $a\gg c$ the opposite pertains: the fast wave is magnetic and the slow wave is acoustic. In these asymptotic regimes, the fast and slow magneto\-acoustic waves are completely decoupled and distinguishable. However, in the conversion region ($a\approx c$) they interact. Importantly, the ``fast'' and ``slow'' waves do not maintain their identities through this coupling region.

Let us be clear about the meaning of ``transmission'' and ``conversion'' here. We say that a wave is totally transmitted ($T=1$) if it maintains its acoustic or magnetic identity across the conversion layer. For example, a vertically propagating sound wave in vertical magnetic field does not interact with the magnetic field, since it is longitudinal, and stays a sound wave as it passes through $z_\mathrm{eq}$: we call this total transmission. Conversely, an incident acoustic wave in $c\gg a$ which passes through to become purely a magnetic wave is said to have been totally converted. Of course, in general, both transmission and conversion are partial.

GRT is only an approximate description of wave behaviour. It shares many of the flaws of standard MHD ray theory \cite{w62}, though not the gross error of presuming perfect fast-fast or slow-slow connectivity across $z_\mathrm{eq}$ (\emph{i.e.}, $T=0$). Specifically, it should be asymptotically correct in the high frequency limit, where wavelengths are vanishingly small compared to background inhomogeneity length scales. But how well does it perform in more realistic, moderate-frequency scenarios? In this paper, we test the accuracy of the GRT estimation of transmission coefficient $T$ in a simple, uniform-field, isothermal, gravitationally-stratified model for which exact wave solutions exist. Although not exhibiting all of the features that we may wish in a sunspot model, it does possess the most important characteristic -- a rapidly increasing $a/c$ ratio with height -- which allows for a meaningful and informative test.

\section{Model and Equations}   \label{sec:equations}
We adopt the simplest model which exhibits the required features of increasing $a/c$ with height, with $a=c$ at some finite level: an isothermal, gravitationally-stratified atmosphere with uniform magnetic field inclined at angle $\theta$ to the vertical. Without loss of generality, the equipartion level ($z_\mathrm{eq}$) is set at zero. A sound wave with horizontal and temporal dependence $\exp[\ri(k_x\,x-\omega\,t)]$ is launched upward from great depth, is incident on $z=0$ where it is partially transmitted and partially converted, and the resulting transmitted sound wave is examined in $z>0$. The aim is to calculate the transmission coefficient ($T$) at varying frequencies ($\omega$), horizontal wavenumbers ($k_x$), and field inclinations ($\theta$). In order that the sound wave is vertically propagating rather than evanescent, and that it is indeed an acoustic and not a gravity wave, $(\omega,k_x)$ is chosen to lie in Region I of the acoustic-gravity wave dispersion diagram \cite[Figure 1]{cally01}, \emph{i.e.}, above the upper branch of $\omega^2-\omega_c^2-c^2 k_x^2(1-N^2/\omega^2)=0$, where $\omega_c=c/2H$ is the acoustic cut-off frequency, $N^2=g/H-g^2/c^2$ is the square of the {\bv} frequency, $g$ is the gravitational acceleration, and $H$ is the density scale height. This requires $\omega>\omega_c$ for $k_x=0$, and a slightly more stringent condition for nonzero $k_x$.
However, in strong, inclined magnetic field, the ``ramp effect'' reduces the effective cut-off frequency to $\omega_c\cos\theta$ \cite{scb84,sc06,mj06}. This effective cut-off term will become apparent in the exact solutions of Section \ref{sec:exact}.

\subsection{Exact Solution}  \label{sec:exact}
Following \citeauthor{cally01} (\citeyear{cally01}, \citeyear{cally08}), we define the following dimensionless variables: frequency $\nu=\omega H/c$, vertical position $\zeta=\omega H/a$ where $H$ is the (constant) density scale height, and the {\bv} frequency $n=NH/c=(\gamma-1)^{1/2}/\gamma$, where $\gamma$ is the usual ratio of specific heats. In these units, the acoustic cut-off frequency is $\half$. Since $a=c\, \mathrm{e}^{z/2H}$ it follows that $\zeta=\nu\, \mathrm{e}^{-z/2H}$ and that $\zeta\rightarrow0^{+}$ as $z\rightarrow\infty$. Finally, $\kappa=k_x\,H$ is the dimensionless horizontal wavenumber, $\kappa_{0}=\sqrt{\nu^2\sectt-1/4}\,$, and $\kappa_z =  \sqrt{\nu^2+(n^2-\nu^2)\kappa^2/\nu^2-1/4}$ is the vertical dimensionless wavenumber in the equivalent non-magnetic case. The definition of $\kappa_0$ directly invokes the ramp effect, as it is real only for $\omega\ge \omega_c^2\cos^2\theta$.

\begin{table}  \label{atable}
\caption{The $a$-coefficients which appear in equation (\ref{eq:transexact}) for the exact transmission coefficient $\mathcal{T}$.}
\begin{tabular}{lc}
\hline\\
 $a_{2\,2}$ & $\frac{\sqrt[4]{-1} e^{\pi \kappa (\tan{\theta}+\ri)} \kappa \Gamma (2 \kappa ) \Gamma
 \left(-\ri \kappa\tan{\theta} +\kappa -\ri \kappa _0+\frac{1}{2}\right) \Gamma \left(-\ri \kappa\tan{\theta} +\kappa +\ri \kappa _0+\frac{1}{2}\right) \sec ^{2 \ri \kappa\tan{\theta} -2
 \kappa -\frac{1}{2}}{\theta}}{\sqrt{\pi } \Gamma \left(\kappa -\ri \kappa
 _z+\frac{1}{2}\right) \Gamma \left(\kappa +\ri \kappa _z+\frac{1}{2}\right)}$ \\[10pt]
 $a_{2\,3}$ & 
$\frac{\sqrt[4]{-1} e^{\pi \kappa _0} \Gamma \left(-2 \ri \kappa _0\right) \Gamma \left(\ri \kappa\tan{\theta} -\kappa -\ri \kappa _0+\frac{3}{2}\right) \Gamma \left(\ri \kappa\tan{\theta} +\kappa -\ri \kappa _0+\frac{3}{2}\right) \kappa _0\sec ^{2 \ri \kappa _0-\frac{3}{2}}{\theta}
 }{\sqrt{\pi } \Gamma \left(-\ri \left(\kappa _0+\kappa _z-\kappa \tan\theta\right)+1\right) \Gamma \left(-\ri \kappa _0+\ri \kappa _z+\ri \kappa \tan{\theta}+1\right)}$\\[10pt]
$a_{4\,2}$ & $\frac{\Gamma (2 \kappa +1) \Gamma \left(-2 \ri \kappa _z\right) \Gamma \left(-\ri \tan{\theta}
 \kappa +\kappa -\ri \kappa _0+\frac{1}{2}\right) \Gamma \left(-\ri \kappa\tan{\theta} +\kappa
 +\ri \kappa _0+\frac{1}{2}\right) \sec ^{-2 \kappa -2 \ri \kappa _z-1}{\theta}}{\Gamma
 \left(\kappa -\ri \kappa _z+\frac{1}{2}\right){}^2 \Gamma \left(-\ri \left(-\kappa _0+\kappa
 _z+\kappa \tan{\theta}\right)\right) \Gamma \left(-\ri \left(\kappa _0+\kappa _z+\kappa 
 \tan{\theta}\right)\right)} $ \\[10pt]
$a_{4\,3}$ & $\frac{\Gamma \left(1-2 \ri \kappa _0\right) \Gamma \left(-2 \ri \kappa _z\right) \Gamma \left(\ri
 \kappa\tan{\theta} -\kappa -\ri \kappa _0+\frac{3}{2}\right) \Gamma \left(\ri \kappa\tan{\theta} +\kappa -\ri \kappa _0+\frac{3}{2}\right) \sec ^{2 \ri \kappa _0-2 \ri \kappa _z-2 \ri \kappa
 \tan{\theta}-2}{\theta}}{\Gamma \left(-\kappa -\ri \kappa _z+\frac{1}{2}\right) \Gamma
 \left(\kappa -\ri \kappa _z+\frac{1}{2}\right) \Gamma \left(-\ri \left(\kappa _0+\kappa
 _z-\kappa \tan{\theta}\right)+1\right) \Gamma \left(-\ri \left(\kappa _0+\kappa _z+\kappa 
 \tan{\theta}\right)\right)}$ \\[10pt]
\hline
\end{tabular}
\end{table}

With these variables, the linearized MHD wave equations may be expressed as a single fourth order ordinary differential equation of hypergeometric type for $u$, the component of velocity perpendicular to the field. The resulting exact magneto-acoustic-gravity wave solutions may be expressed in terms of Meijer-G functions \cite{zd84}, or more simply the $\F$ hypergeometric function \cite{cally08}. The $\F$ function is extensively discussed in \inlinecite{luke}.

In the degenerate case of horizontal magnetic field, the equation reduces to second order, and the well-known spectral structure discussed at length in Chapter 7 of \inlinecite{gp} results, including cusp (often called slow) and Alfv\'en continua. These continua are due to the coefficient of the highest (\emph{i.e.}, second) derivative vanishing at Alfv\'en and cusp ``critical levels'' (see \opencite{cally84} and \opencite{gp}, Section 7.3.2). There are no such critical levels when the magnetic field has a vertical component, and they will not be discussed further.

As set out by \inlinecite{cally08}, the solution for $u$ is 
\begin{multline}
u = \\
C_1\, \zeta^{-2\kappa}\, 
\F\bigl(\ts\half - \kappa - \ri \kpz,\, \half - \kappa + \ri \kpz;
\quad\qquad\qquad\qquad\qquad\qquad\qquad\qquad\qquad\qquad\\\qquad\qquad
 \, 1-2\kappa,\, \half - \kappa - \ri\kp-\ri\kappa\tan\theta,\, \half - \kappa + \ri\kp-\ri\kappa\tan\theta;
\, -\zeta^2\sec^2\theta\bigr) +
\\[6pt]
C_2\, \zeta^{2\kappa} \, 
\F\bigl(\ts\half + \kappa - \ri \kpz,\, \half + \kappa + \ri \kpz;
\quad\qquad\qquad\qquad\qquad\qquad\qquad\qquad\qquad\qquad\\\qquad\qquad
\, 1+2\kappa,\, \half + \kappa - \ri\kp-i\kappa\tan\theta,\, \half + \kappa + \ri\kp-\ri\kappa\tan\theta;
\, -\zeta^2\sec^2\theta\bigr) +
\\[6pt]
C_3\, \zeta^{1-2\ri\kp+2\ri\kappa\tan\theta} \, 
\F\bigl(\ts1-\ri\kp-\ri \kpz+\ri\kappa\tan\theta,\,1-\ri\kp+\ri \kpz+\ri\kappa\tan\theta;
\qquad\qquad\\\qquad\qquad
\,1-2i\kp,\,\threeontwo-\ri\kp-\kappa+\ri\kappa\tan\theta,\,\threeontwo-\ri\kp+\kappa+\ri\kappa\tan\theta;
\,-\zeta^2\sec^2\theta\bigr) +
\\[6pt]
C_4\, \zeta^{1+2\ri\kp+2\ri\kappa\tan\theta} \, 
\F\bigl(\ts1+\ri\kp-\ri \kpz+i\kappa\tan\theta,\,1+\ri\kp+\ri \kpz+\ri\kappa\tan\theta;
\qquad\qquad\\\qquad\qquad
\,1+2\ri\kp,\,\threeontwo+\ri\kp-\kappa+\ri\kappa\tan\theta,\,\threeontwo+\ri\kp+\kappa+\ri\kappa\tan\theta;
\,-\zeta^2\sec^2\theta\bigr)\, ,
                                                           \label{u}
\end{multline}
where the $C_i$ are the arbitrary constant coefficients. The individual solutions $u_1$, \ldots, $u_4$ multiplied by the $C$-coefficients are said to be of Types 1 to 4 respectively.  Since the $\F$ functions all approach 1 as $\zeta\to0$ ($z\to\infty$), the asymptotic behaviour of these four solutions is evident, and they may be identified as the exponentially growing fast mode, the evanescent fast mode, the outgoing slow (acoustic) mode, and the incoming slow mode respectively. Applying top regularity and radiation boundary conditions, the first and fourth solutions are therefore dropped, so the physical solutions are necessarily a linear combination of $u_2$ and $u_3$. Similarly, we may apply a bottom boundary condition ($z\to-\infty$, \emph{i.e.}, $\zeta\to\infty$) that there be no incoming slow modes (allowing only an incoming fast mode and outgoing slow and fast waves). \inlinecite{cally08} shows that the physical solution for this case of an incident acoustic (fast) wave from below is $a_{2\,3}u_2-a_{2\,2}u_3$, where the $a$-coefficients are expressed in terms of $\Gamma$-functions (Table~\ref{atable}). 

The total wave energy flux and direction can be established for each wave type, with the vertical flux of each wave mode remaining constant with height \cite{zd84}.  The incident fast wave is purely acoustic at large height, so neglecting the magnetic term the vertical flux is
\begin{equation}
F_{z}  = \re[p_1\,v_z^*]= \re\left[\ri\,\frac{\zeta}{\nu^2}\left(-\ri\,\kappa\,\xi_x+\frac{\zeta}{2}\,\xi_z^{\prime}\right)\xi_z^{*}\right]\, ,
\label{verticalflux}
\end{equation} 
where $p_1$ is the Eulerian pressure perturbation, $v_z$ is the $z$-component of the plasma velocity, and $\xi_x$ and $\xi_z$ are defined as the horizontal and vertical components of displacement respectively.

The exact transmission coefficient ($\mathcal{T}$) is defined as the proportion of incident wave energy flux transmitted from fast to slow acoustic waves.  It takes a similar form to that found in the vertical field case by \inlinecite{cally01}, and with the appropriate $a$-coefficients is 
\begin{equation}
\mathcal{T}  = \frac{|a_{2\,2}|^2\,\phi}{|a_{2\,3}a_{4\,2}-a_{2\,2}a_{4\,3}|^2\,f}\,.
\label{eq:transexact}
\end{equation}
Here the incident fast flux is
\begin{multline}
f  =  b_2 \Biggl\{ \kappa^2\kappa_z\Biggl[4(\gamma-1)\kappa^2\nu^2+4n^2\biggl(\kappa^2\Bigl(1+\gamma(n^2\gamma-1-\gamma\nu^2)\Bigr)\\
-(\gamma-1)\kappa\nu^2\Bigl(-4n^2\kappa^2+\nu^2(1+4\kappa^2-4\nu^2)\sin2\theta\Bigr)\\
+\nu^2\Bigl(1+\gamma(\gamma\nu^2-1)\Bigr)\biggr)\Biggr]\costt+4(\gamma-1)\kappa_z\nu^4(\nu^2-\kappa^2)\sint\Biggr\}\, ,
\label{eq:f}
\end{multline}
where 
\begin{multline*}
\frac{1}{b_2}  =  \nu\Biggl[(\gamma-1)\Bigl(n^2\kappa^2-2\kappa^2\nu^2+\nu^4+(n^2\kappa^2-\nu^4)\cos2\theta\Bigr)^2\\
+n^2(\gamma-2)^2\kappa^2\nu^4\sin^22\theta\biggr] \,.       
\end{multline*}
Similarly the vertical transmitted flux is
\begin{multline}
\phi  = b_3\Biggl\{\Biggl[\gamma^2\nu^4+\kappa^2\Bigl(1+\gamma(2\gamma\nu^2-1)\Bigr)-\gamma^2\Bigl((n^2\kappa^2+\nu^4)\cos2\theta-2\kappa\kappa_0^2\nu^2\sin2\theta\Bigr)\Biggr]\\
\times \kappa_0\nu\Biggl[48\ka_0^4-3+64\Bigl(2\ka^4+2\nu^4+\ka^2(8\nu^2-1)\Bigr)-8\biggl(8\ka^2(1+4\nu^2)\cos2\theta\\
-3\nu^2\sectt+6\nu^4\sec^4\theta+4\ka\ka_0\Bigl(1+8\ka^2+4\kappa_0^2+(1+4\kappa_0^2)\cos2\theta\Bigr)\sin2\theta\biggr)\Biggr]\Biggr\}\, ,
\label{eq:phi}
\end{multline}
where
\begin{equation*}
\frac{1}{b_3}  =  \Biggl[64\biggl(\gamma^2\Bigl(n^2\ka^2-2\kappa\nu^2+\nu^4+(n^2\kappa^2-\nu^4)\cos2\theta\Bigr)^2+(\gamma-2)^2\kappa^2\nu^4\sin^22\theta\biggr)\Biggr]\, .
\end{equation*}

\subsection{Ray Solution}  \label{sec:ray}

Standard ray theory treats waves like a particle moving in phase space.  It is applicable in weakly inhomogeneous media in the high frequency limit.  The 2D magneto\-acoustic-gravity wave dispersion function on which it may be based is
\begin{equation}
\calD = \calD_{2\mathrm{D}}=\omega^4 - (a^2+c^2)k^2\omega^2 + a^2 c^2
k^2\kpar^2+c^2 N^2 k_x^2 - (\omega^2-a^2 k^2\cos^2\theta)\omega_{c}^2\, , 
\label{D}
\end{equation}
where $k=(k_x^2+k_z^2)^{1/2}=|\k|$ is the magnitude of the wavevector, $k_x$ and $k_z$ are its horizontal and vertical components, and $\kpar$ its component along the magnetic field direction.
Locally, the dispersion relation $\mathcal{D}=0$ restricts the solutions within frequency-wavevector phase space to fast and slow hypersurfaces \cite{sc06}.  The ray paths are derived from the Hamiltonian equations 
\begin{equation}
{\rd {\bf x} \over \rd\tau} = {\partial\mathcal{D} \over \partial {\bf k}}, 
 \quad
{\rd {\bf {\bf k}} \over \rd\tau} = -{\partial\mathcal{D} \over \partial {\bf x}},
\quad
{\rd t \over \rd \tau} = - {\partial\mathcal{D} \over \partial \omega},
\quad
{\rd\omega \over \rd \tau} = {\partial\mathcal{D} \over \partial t}, \label{ray}
\end{equation}
where $\tau$ is a time-like parameter following the disturbance. In the present case, the dispersion relation is independent of horizontal position, ($x$), and time ($t$), meaning that frequency ($\omega$) and horizontal wavenumber ($k_x$) are constant \cite{w62}. The inclusion of acoustic cut-off and {\bv} frequencies means that the ray paths no longer obey Fermat's principle \cite{bc01}.

Using the foundations set out by \inlinecite{w62} and  \inlinecite{tkb03}, \inlinecite{cally06} and \inlinecite{sc06} developed GRT to allow for mode conversion within the ray description. Standard ray theory breaks down at close avoided crossings of the fast and slow phase loci in $z$-$k_z$ phase space, but GRT redresses this by effectively doing a wave-mechanical matching across these regions. This process then returns the (approximately) correct connectivity. Readers are referred to \inlinecite{sc06} for a full description. We simply precis the mathematical process here, without further elaborating on its foundations. 

The mode transmission/conversion capabilities of GRT are based on constructing the dispersion function 
\begin{equation}
\mathcal{D}  =  \det\mathbf{D} = D_aD_b-|\eta|^2
\end{equation} 
from a Hermitian dispersion matrix
\begin{equation} \mathbf{D}
 = \left(            \begin{array}{ll}   
                 D_a & \eta \\                   
                 \eta^{*} & D_b              
                 \end{array}       
                 \right) \, .
\end{equation}
Specific expressions for $D_a$ and $D_b$ are given in \inlinecite{sc06}.
Now, $\eta$ is ``small'' away from the conversion/transmission region, and so the dispersion relation $0=\mathcal{D}\approx D_a\,D_b$ yields $D_a=0$ or $D_b=0$, the decoupled acoustic and magnetic waves. However, close to the ``star point'' where $D_a=D_b=0$, the coupling term $\eta$ dominates, and provides the local connectivity description. The fraction of energy transmitted from acoustic-to-acoustic or magnetic-to-magnetic is given by
\begin{equation}
T  =  T_\mathrm{f}=\exp\left[-2\pi|\eta|^2/|{\cal B}|\right]_{*}\, ,
\label{transfull}
\end{equation}
where 
\begin{equation}
\mathcal{B}=\{D_a,D_b\}=\frac{\partial D_a}{\partial k_z}\frac{\partial D_b}{\partial z}-\frac{\partial D_b}{\partial k_z}\frac{\partial D_a}{\partial z}
\end{equation}
is the Poisson bracket of the uncoupled dispersion functions. The subscripted star indicates that the expression should be evaluated at the star point where $D_a$ and $D_b$ simultaneously vanish. Hence
$|\eta|_{*}^2 =  -\mathcal{D}_* $. Clearly, $T=1$ if $\eta_*=0$, representing total transmission. The subscript ``f'' in $T_\mathrm{f}$ is used here to distinguish the ``full'' form of the transmission coefficient from the ``simplified'' form, which we now introduce.

\begin{figure}
\begin{center}
\includegraphics[width=\hsize]{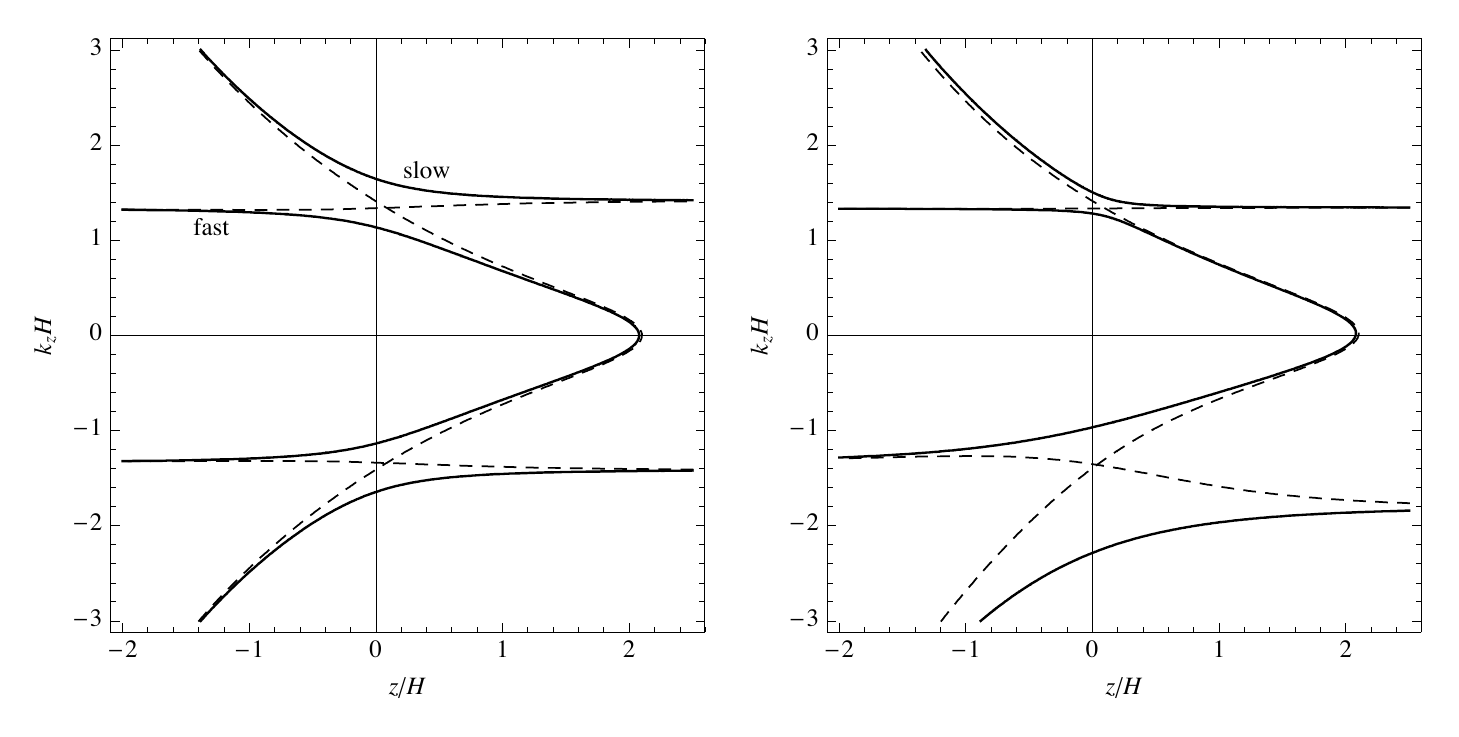}
\caption{Phase space diagrams ($z$-$k_z$) for the case $\kappa=0.5$, $\nu=1.5$, $\gamma=5/3$. Left: $\theta=0^\circ$; right: $\theta=25^\circ$. The fast and slow loci (full curves) are labelled in the left panel. The $D_a=0$ and $D_b=0$ loci are represented as dashed curves. The star points are where these cross. In the left panel $T_\mathrm{f}=0.63$ at both star points (the vertical field case is always symmetric), whilst in the right panel $T_\mathrm{f}=0.93$ for the close avoided crossing and $T_\mathrm{f}=0.19$ for the wide one.}
\label{avoid}
\end{center}
\end{figure}

Figure \ref{avoid} displays the avoided crossings in $z$-$k_z$ phases space for typical vertical and inclined field cases, and clearly illustrates how magnetic field inclination can lead to much higher (and lower) transmission coefficients by narrowing (or widening) the gap between the fast and slow loci.

At high frequency, $\omega_c$ and $N$ can be neglected, moving the star point exactly to $a=c$, and yielding a simplified and intuitively useful expression for the transmission:
\begin{equation}
T_\mathrm{s} = \exp\left[-\frac{\pi\, h\, k^2 \kperp^2}
{|k_z|\,(k^2+\kperp^2)}\right]_{a=c}\, ,    \label{Ts}
\end{equation}
where $h=[\rd(a^2/c^2)/\rd z]^{-1}_{a=c}$ is the equipartition layer scale height. Note the dependence on the perpendicular component of the wavevector $\kperp=k\sin\alpha$: when the wavevector is parallel to the field it vanishes, and transmission becomes total in the simplified description. This is not exactly the case for $T_\mathrm{f}$. In the current isothermal context, Equation (\ref{Ts}) takes the form 
\begin{equation}
T_\mathrm{s} =\exp\left[-\frac{\pi\,  \nu ^2 \left(\kappa  \cos \theta -\sqrt{\nu ^2-\kappa ^2}\, \sin \theta
   \right)^2}{\sqrt{\nu ^2-\kappa ^2} \left(\nu ^2+\left(\kappa  \cos \theta -\sqrt{\nu
   ^2-\kappa ^2} \,\sin \theta \right)^2\right)}\right]_{a=c}\, .
   \label{simplegenform}
\end{equation}

The full transmission formula for $T_\mathrm{f}$ retains the acoustic cut-off and {\bv} frequencies, and so should be more accurate, although also more complex. Both $T_\mathrm{f}$ and $T_\mathrm{s}$ are tested against the exact transmission $\mathcal{T}$ in Section \ref{sec:results}.

\section{Results}  \label{sec:results}

\begin{figure}
\begin{center}
\includegraphics[width=\textwidth]{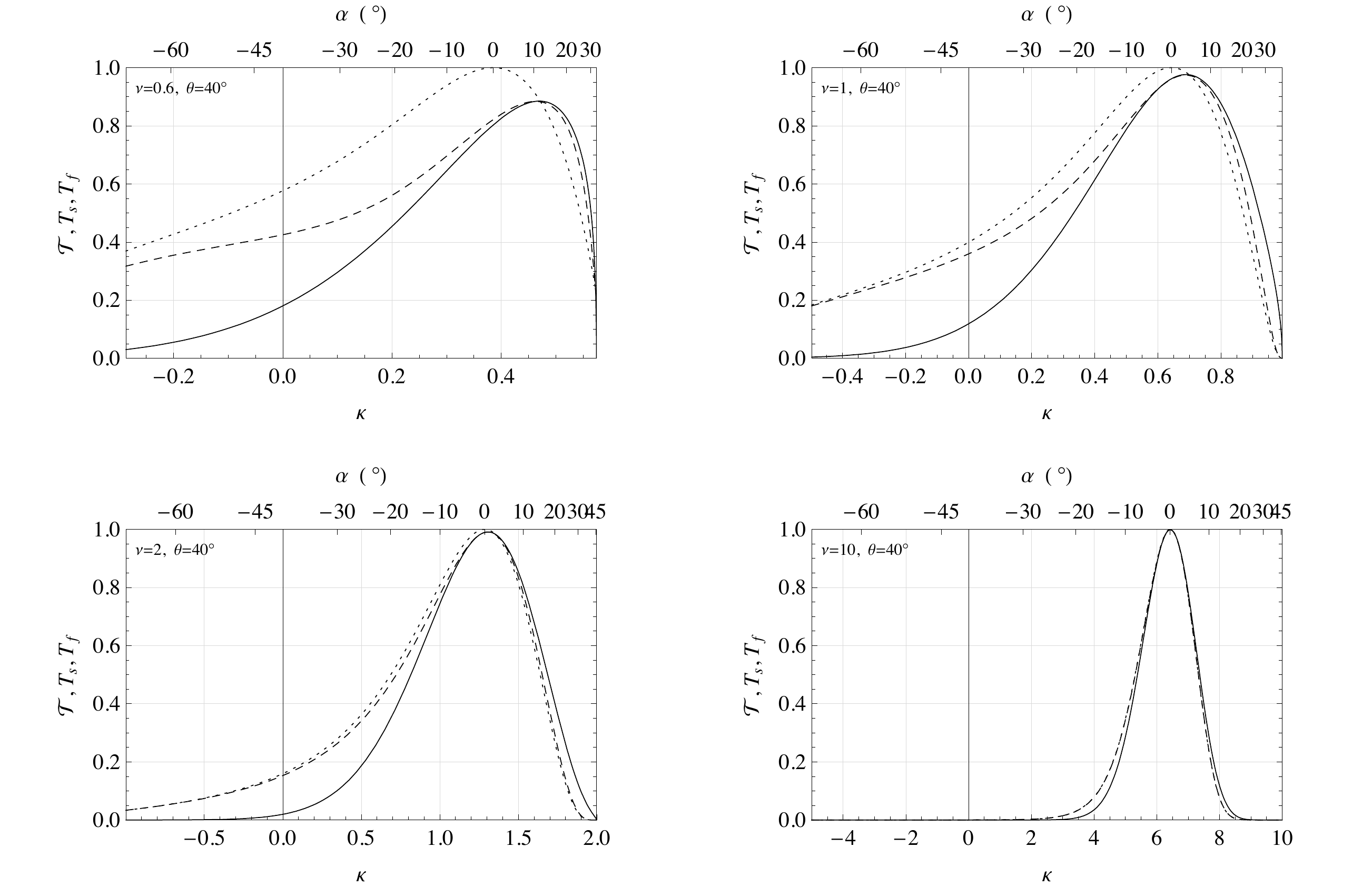}
\caption{ Exact (solid line), full GRT (dashed line), simple GRT (dotted line) transmission formula plotted against $\kappa$ with $\theta=40^{\circ}$ for $\nu=0.6$ (top left), $\nu=1$ (top right), $\nu=2$ (bottom left) and $\nu=10$ (bottom right).  Accuracy of both GRT formulas increases in the high frequency limit.}
\label{frequency}
\end{center}
\end{figure}

\begin{figure}
\begin{center}
\includegraphics[width=1\textwidth]{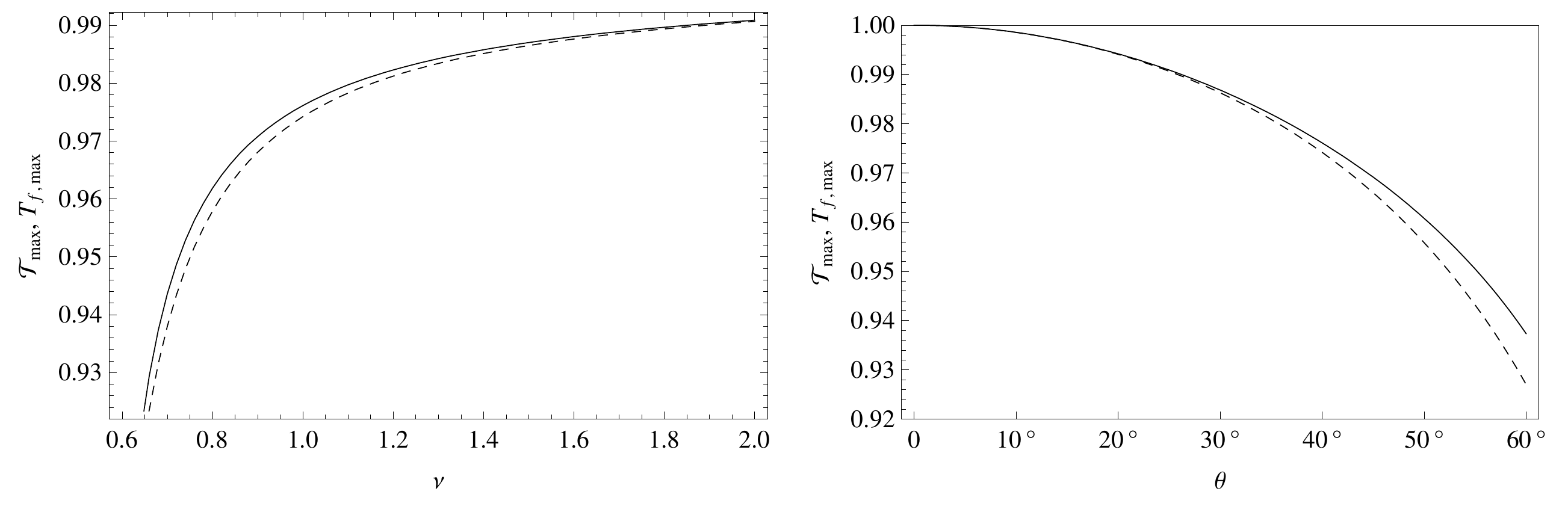}
\caption{Exact (solid line) and full GRT (dashed line) maximum transmission points plotted against $\nu$ for $\theta=40^{\circ}$ (left) and the magnetic field inclination angle for $\nu=1$ (right).}
\label{maxtransmission}
\end{center}
\end{figure}

\begin{figure}
\begin{center}
\includegraphics[width=1\textwidth]{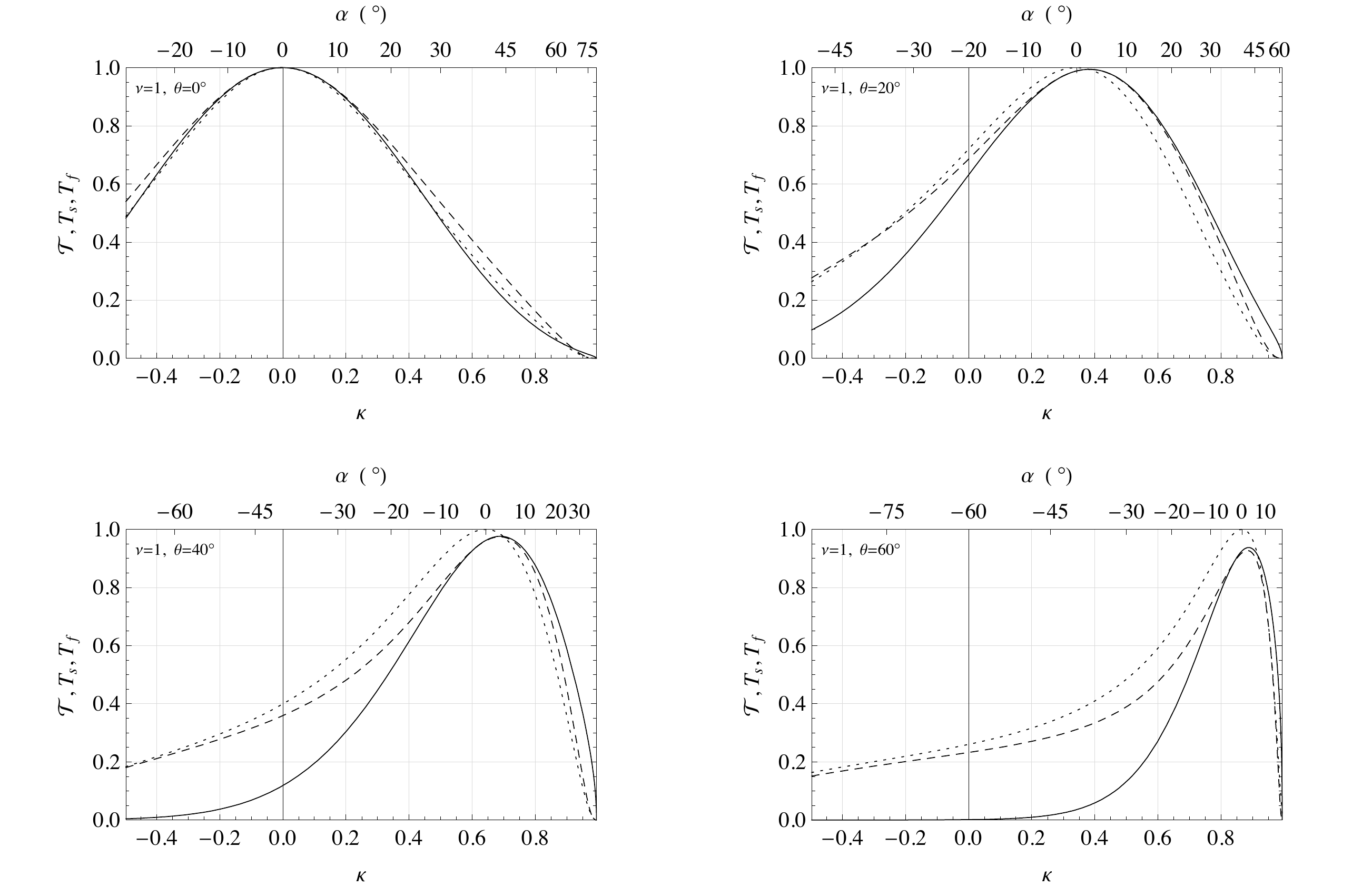}
\caption{Exact (solid line), full GRT (dashed line), simple GRT (dotted line) transmission formula plotted against $\kappa$ with $\nu=1$ for $\theta=0$ (top left), $\theta=20^{\circ}$ (bottom left), $\theta=40^{\circ}$ (top right) and $\theta=60^{\circ}$ (bottom right).  Accuracy of both GRT formulas decreases with high magnetic field inclinations.}
\label{inclination}
\end{center}
\end{figure}

Figure \ref{frequency} displays $\mathcal{T}$, $T_\mathrm{f}$, and $T_\mathrm{s}$ as functions of dimensionless horizontal wavenumber ($\kappa$) for magnetic field inclined at $40^\circ$ to the vertical and various frequencies $\nu$ ranging from only slightly above the acoustic cut-off at $\nu_c=\half$ and {\bv} frequency $n=0.4899$, where $\gamma=\frac{5}{3}$ is used throughout, up to $\nu=10=20\nu_c$. As $\kappa$ increases, so does the inclination of the wavevector; at a particular value it is parallel to the magnetic field, resulting in $\kappaperp=\kperp \, H$ vanishing and $T_\mathrm{s}$ peaking at 1. The attack angle ($\alpha$) is also displayed on the top axis. (The upper limit attained by $\kappa$ in these graphs is due to the model leaving Region I of the acoustic-gravity propagation diagram beyond that point.)

As expected, both approximate formulae perform very well at high frequency. The simple ray formula of course incorrectly returns a peak transmission of 1 at $\alpha=0$, but remarkably the full ray formula manages to almost perfectly fit the exact peak, even at very low frequency. Figure \ref{maxtransmission} shows just how well the peak transmission is approximated by $T_\mathrm{f}$ for a range of frequencies and magnetic field inclinations. It is not surprising that even $T_\mathrm{f}$ fails at low frequency and large attack angle, as the foundational transmission formula (\ref{transfull}) is derived from a local analysis around the star point that becomes progressively more inaccurate as the avoided crossing widens (attack angle increases). 

Figure \ref{inclination} similarly fixes frequency at $\nu=1$ and plots $\mathcal{T}$, $T_\mathrm{f}$, and $T_\mathrm{s}$ against $\kappa$ (or $\alpha$) for various field inclinations. Both $T_\mathrm{f}$ and $T_\mathrm{s}$ cope with near-vertical field better than highly inclined field, although the peak is well-fitted by the full ray formula throughout.

As an aside, we mention that the exact transmission curves presented here are not consistent with Figure 2 of \inlinecite{zd84b}, where transmission coefficient peaks are typically much lower. However, the beautiful correspondence between the completely independent exact and full GRT solutions here provides confirmation of these present results.

\section{Conclusion}  \label{sec:conclusion}
Generalized Ray Theory (GRT) redresses one of the major failings of classical ray theory in the MHD context; the failure to recognize wave transmission/conversion at avoided crossings of the fast and slow loci in phase space. It does this by adding a local wave-mechanical analysis in the neighbourhood of these ``star points'' in a way that is quite convenient for practical calculation. The applications to helioseismology by \inlinecite{sc06} and \inlinecite{cally07} have provided important new insights into the physics of near-surface mode mixing, which complement lessons learned from direct numerical solution of the wave equations (\emph{e.g.}, \opencite{cally00}; \opencite{kc06}).

However, it is important to realize the strengths and limitations of this new tool. As with all ray theory, it works best at small wavelength. However, we have seen here that it is nonetheless extremely useful even at frequencies comparable to $\omega_c$ and $N$. The nature of the isothermal model precluded our reducing the frequency further, below $\omega_c$, as the required incident sound wave then does not propagate at all.

It is important to appreciate though that the true value of GRT lies not in its ability to yield accurate transmission coefficients, which it certainly can do near the transmission peaks, but rather in its qualitative and physically compelling description of the processes in play.

\begin{acknowledgements}
\end{acknowledgements}



\end{document}